\documentclass[
reprint,
amsmath,amssymb,prl,
aps,superscriptaddress]{revtex4-2}

\usepackage{graphicx}
\usepackage{dcolumn}
\usepackage{bm}
\usepackage[ruled]{algorithm2e}
\usepackage{multirow,booktabs}
\usepackage{svg}
\usepackage{paralist,siunitx,soul}
\usepackage[version=4]{mhchem}
\usepackage{hyperref}
\usepackage{url}
\hypersetup{colorlinks,urlcolor=blue,citecolor=blue,linkcolor=blue}
\DeclareUrlCommand\url{\color{blue}}

\begin{document}
	
\title{Revisiting the theory of crystal polarization: The downside of employing  the periodic boundary conditions}
\author{Qiu-Shi Huang}
\affiliation{Eastern Institute of Technology, Ningbo 315200, China}
\affiliation{University of California, Santa Barbara, California 93106-5050, United States}

\author{Su-Huai Wei}
\email{suhuaiwei@eitech.edu.cn}
\affiliation{Eastern Institute of Technology, Ningbo 315200, China}

\date{\today}

\begin{abstract}
	Periodic boundary condition (PBC) is a standard approximation for calculating crystalline materials properties. However, a PBC crystal is not the same as the real macroscopic crystal, therefore, if applied indiscriminately, it can lead to erroneous conclusions. For example, unlike other extensive observables such as total energy, the polarization of a macroscopic crystal, cannot always be described by a PBC model, because polarization is inherently nonlocal and strongly dependent on surface terminations, irrespective of crystal size, and moreover, the symmetry of the macroscopic crystal can be altered when the PBC is applied to a macroscopic crystal. We demonstrate in this paper that the polarization of a macroscopic crystal receives contributions from both the repeating bulk units and the crystal surfaces, which must be treated on an equal footing. When the combined system of the bulk and its surfaces are taken into account, materials traditionally classified as nonpolar can, in fact, admit polar symmetry, thus explaining why experimentalists have observed polarization in some nominally “nonpolar” systems. Our study, thus, clarifies that polarization can only exist in polar group systems and that apparent violations of the Neumann’s principle reported in some recent works originate from misinterpreting bulk PBC crystal as intrinsic macroscopic crystal, ignoring the contribution from the surfaces. We demonstrate that when the full bulk-plus-surface system is considered, the crystal polarization and symmetry is fully consistent with Neumann’s principle. 
\end{abstract}

\maketitle

The concept of electric dipole moment plays a crucial role in electrostatics, especially in understanding a system's response to electric fields. It is a textbook knowledge that the electric dipole moment can only exist along polar directions in a polar crystal with the following ten point groups: \(C_{6v}\), \(C_6\), \(C_{4v}\), \(C_4\), \(C_{3v}\), \(C_3\), \(C_{2v}\), \(C_2\), as well as \(C_s\) and \(C_1\). For the first eight groups, the electric dipole moment can only exist along the rotation axis\cite{PolarMetals}. In the \(C_s\) group, the polarization direction can only be parallel to the mirror plane, while in the \(C_1\) group, due to the absence of any symmetry, the electric dipole moment can occur along any direction. Traditional ferroelectricity is often associated with a structural phase transition associated with atomic displacements from an inversion-symmetric non-polar structure to a noncentrosymmetric polar structure (the symmetry of the structure belongs to the ten polar crystal point groups). However, in recent years, some researchers have claimed the discovery of many new ferroelectric materials that do not belong to the ten polar point groups\cite{yang2022ferroelectricity, seleznev2023cyclic, ji2024fractional,cordero2019flexoelectric, hu2022bond, qi2021widespread, ding2021two, qi2020stabilization, chai2020nonvolatile, ma2023ultrahigh, zhou2020anomalous, wang2022unconventional, gao2019phase}. Typically, periodic boundary condition (PBC) is assumed to explain the crystal symmetry and modern theory of polarization (MTP) \cite{Spaldin2012, baroni1987, kingsmith1993, resta1993, vanderbilt1993, resta1998, resta1994} is used to calculate polarization of these new ferroelectric materials.

The MTP has undoubtedly been very successful in calculating ferroelectric polarization. Starting from wave functions obtained from first-principles calculations, it rigorously and compactly evaluates the polarization change associated with ferroelectric switching by integrating the Berry phase along a fixed path. Unfortunately, due to their success in practical computations, people have gradually begun to blur the distinction between the properties of PBC systems and those of real systems. As a result, conclusions obtained strictly within a periodic picture are at times treated as intrinsic crystal properties, fostering misconceptions (e.g., that the polarization in a real macroscopic systems is “multivalued,” as in a PBC system). Moreover, several recent studies have been interpreted as observing finite polarization in nonpolar systems  \cite{ji2024fractional,XiangPhysRevLett.134.016801,trhd-kxm1,nanolett.7b,acsnano.8b,PhysRevLett.120.227601}, thus regarded by many researchers as a violation of the Neumann’s principle. However, once the actual surfaces of real crystals are specified, the symmetry of the relevant bulk-plus-surface system is fully consistent with the Neumann’s principle \cite{neumann1885vorlesungen}, thereby resolving these apparent contradictions.

\begin{figure}[!t]
	\includegraphics[width=8cm]{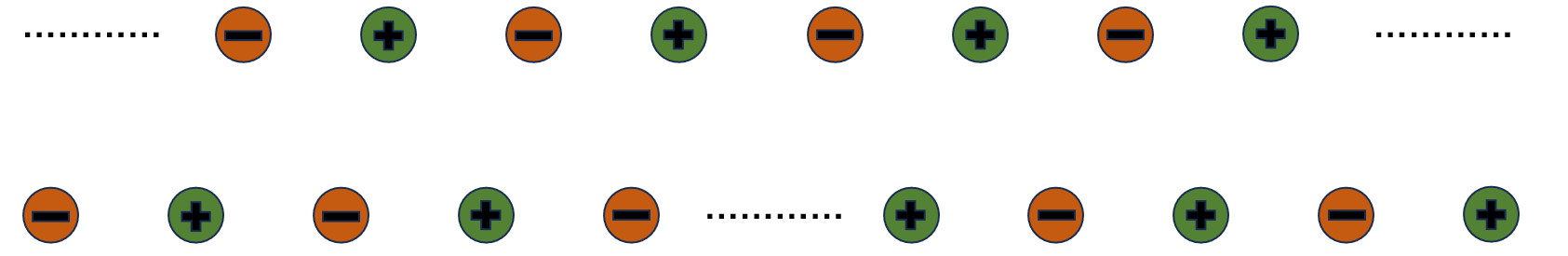}
	\caption{\label{fig:0}  The difference between an periodic crystal (top) and a real infinitely large  crystal (bottom), illustrated using a one-dimensional charged chain as an example, is that in terms of symmetry, the periodic crystal belongs to the nonpolar symmetry group \( D_{\infty} \), whereas the actual real crystal effectively exhibits polar symmetry group \( C_{\infty} \).
	}
\end{figure}
The contravecy can be understood as follows. Due to the PBC imposed, a macroscopic crystal is mimicked by a PERIODIC crystal, in which the two ends are connected(top of Fig.~\ref{fig:0}). However, real macroscopic crystalline materials are inevitably of finite size, with well defined surfaces, despite the number of repeating cells of the crystal can approach infinitely large(bottom of Fig.~\ref{fig:0}). These two system are equivalent only if the implicit assumption that---as the number of repeating unit cells in a crystal increases to macroscopic scale, the contribution from the surfaces to the crystal’s propery PER UNIT CELL becomes negligible---is correct.

However, unlike total energy, and other local extensive material properties, the unit polarization of a macroscopic crystal often can not be described by that of a periodic crystal, even the number of repeating cells increases to infinite. This is because the application of PBC neglects the contribution from the surface, but the effect of the surface on the polarization is of the same order of magnitude as the interior repeating unit cell no matter how large the system is. More seriously, applying PBC often gives wrong crystal symmetry, which leads to unphysical conceptions. Therefore, in this paper, we show that in a macroscopic crystal \(C\), the unit polarization \(P\) of the crystal originates from two parts: the unit polarization of the repeating unit cell \( P_U \) and the contribution of the remaining part of \( C \) after removing the repeating unit cells, which we denote as the surface polarization contribution \( P_S \). Our argument demonstrates that for neutral macroscopic solids, the unit polarization \( P \) of a macroscopic crystal is independent of the choice of unit cell, single-valued for a specific surface condition and well-defined quantity. Consequently, \( \Delta P \) is a state function, depending only on the initial and final states. In considering the symmetry of a macroscopic crystal, the symmetry of the repeating unit cell and the surface must also be taken into account together to determine the correct overall symmetry, which is essential for a proper analysis and interpretation of the physical properties. This global (bulk–plus–surface) symmetry also explains why materials traditionally labeled as ``nonpolar" can exhibit an observable, ``unexpected" macroscopic polarization.

In the ionic limit, the dipole moment, $\mathbf{d}$, of a neutral crystal with a collection of charges, $q_i$, at positions $\mathbf{r}_i$ is defined as
\begin{equation}
	\mathbf{d} = \sum_i q_i \mathbf{r}_i,
\end{equation}
for the case of a continuous charge distribution with density, $n(\mathbf{r})$ this expression is straightforwardly extended to
\begin{equation}
	\mathbf{d} = \int n(\mathbf{r}) \mathbf{r} \, d\mathbf{r}. \label{q1}
\end{equation}

For a macroscopic neutral crystal, this is a well-defined single value quantity. In general, for macroscopic crystal \( C \), we can choose a unit cell \( U \) and divide the crystal \( C \) into two parts: a repeating part \( n \cdot U \) composed of \( U \), and the remaining part as surface \( S \).
\begin{align}
	C &= n \cdot U + S, \\
	S &= C~mod~U. \notag
\end{align}
For the per unit polarization \( P \) of a crystal \( C \),
\begin{align}
	\mathbf{P} &= \frac{\mathbf{d}}{V_{tot}} = \frac{\sum^n \mathbf{d_{U}} + \mathbf{d_{S}}}{V_{tot}} 
	\\
	& = \frac{{V_{U}}}{V_{tot}}\sum^n \mathbf{P_{U}} + \frac{\mathbf{d_{S}}}{V_{tot}}  \notag
	\\
	&= \frac{nV_{U}}{V_{tot}}\mathbf{P_{U}} + \mathbf{P_{S}}.   \notag
\end{align}
For macroscopic solids, \( n \to \infty \),
\begin{align}\label{e0}
	\mathbf{P}
	=\mathbf{P_{U}} + \mathbf{P_{S}},
\end{align}
where \( P_{U} = \frac{d_{U}}{V_{U}} \) represents the contribution from the polarization within the unit cell, where \( V_{U} \) denotes the volume of the unit cell. \( P_{S} = \frac{d_{S}}{V_{tot}} \) represents the contribution of polarization from the surface, where \( V_{tot} \) is the total volume of the macroscopic crystal. In the following discussion, for simplicity and better readability, we map the real quantum system onto an effective point-charge model. We emphasize that this point-charge representation is not an excessive simplification of the underlying quantum system; on the contrary, its equivalence to the full description of the real system can be rigorously established via Wannier or Boys orbitals~\cite{vanderbilt1993,JCPresta,RMPBOYS}.

\begin{figure}[!t]
	\includegraphics[width=8cm]{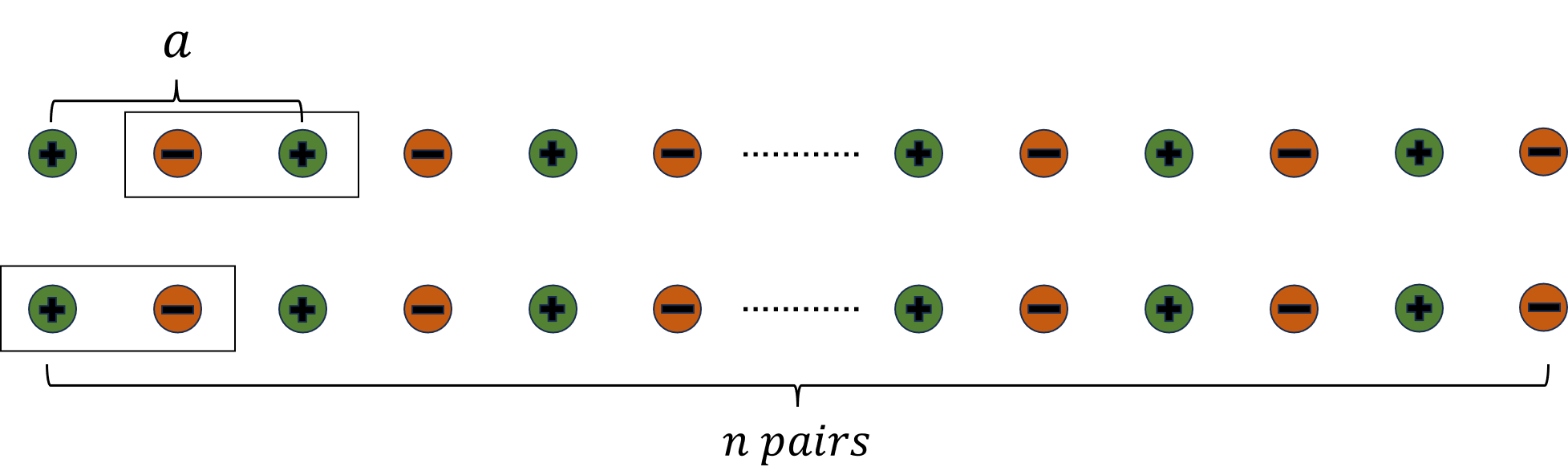}
	\caption{\label{fig:1}  One-dimensional charge chain. The box on the top and the box on the bottom represent different choices of neutral unit cells. (without losing generality, we choose charge \(q=1\) )
	}
\end{figure}
Eq.~\ref{e0} shows that for a real solid \( C \), the polarization \( P \) consists of two parts: \( P_U \) and \( P_S \). Although selecting different \( U \) will lead to different \( P_U \), there will also be different \( P_S \) such that the overall \( P \) of the crystal remains unchanged. Taking the one-dimensional chain shown on Fig.~\ref{fig:1} as an example and choose \(n\) goes to infinity, when we select the unit cell shown on the top in Fig.~\ref{fig:1}, then \( P_U=\frac{1}{2} \) and \( P_S=-1 \), so,
\begin{align}
	\mathbf{P} 
    &=  \mathbf{P_{U}} + \mathbf{P_{S}} 
\\   
    &= \frac{1}{2} -1 \notag 
\\ 
    &= -\frac{1}{2}. \notag 
\end{align}
Note here \( P_S\) is finite and does not go to zero even \(n\) goes to infinity. When we select the unit cell shown on the bottom of Fig.~\ref{fig:1}, then
\begin{align}
	\mathbf{P} 
	&=  \mathbf{P_{U}} + \mathbf{P_{S}} 
	\\   
	&= -\frac{1}{2} + 0 \notag 
	\\ 
	&= -\frac{1}{2}.
\end{align}
So, the polarization is independent of how one chooses the unit cell, as one would expect. Note here \(P = P_U\) only when the unit cell is chosen wisely to make the contribution from the surface goes to zero. However, if we apply PBC to this one-dimensional chain, the chains in Fig.~\ref{fig:1} are forced to change from a polarized chain to a non-polar system with inversion symmetry, and the polarization  becomes multivalued, that is \( P=\frac{1}{2} \) for the top unit cell choice and \( P=-\frac{1}{2} \) for the bottom unit cell choice. In general, under PBC, the polarization \( P \) is given by 
\begin{align}\label{ep}
    \mathbf{P} = \mathbf{P_{U}} + m\mathbf{P_q}.         
\end{align}
Here, \(m\) is an integer. \(Pq\) is the value of polarization resulting from moving one ion with charge \(q\) by one unit lattice vector, i.e., a constant polarization quantum.

This simple example illustrates the essential difference between periodic crystal and macroscopic crystal: the polarization \( P \) of a macroscopic crystal is well-defined, single-valued for a specific surface, and observable. Adding PBC to a macroscopic crystal disregards the surface conditions, but the influence of the surface \( S \) on polarization cannot be ignored.

We want to emphasize here that from a symmetry perspective, as show in Fig.~\ref{fig:0}, a periodic one-dimensional pair chain has center inversion symmetry, and its symmetry group is \(D_{\infty}\), which is a non-polar point group. In contrast, an infinite repeating pair chain has symmetry \(C_{\infty}\), where the rotation axis is the one-dimensional chain, making it a polar point group. The difference in symmetry indicates that symmetry imposed by a periodic boundary condition may not be correct to describe the polarization of a real macroscopic crystals, therefore, one may obtain wrong conclusions by enforcing PBC, such as one can obtain polarization in a "nonpolar" crystal.

\begin{figure}[!t]
	\centering
	\includegraphics[width=8cm]{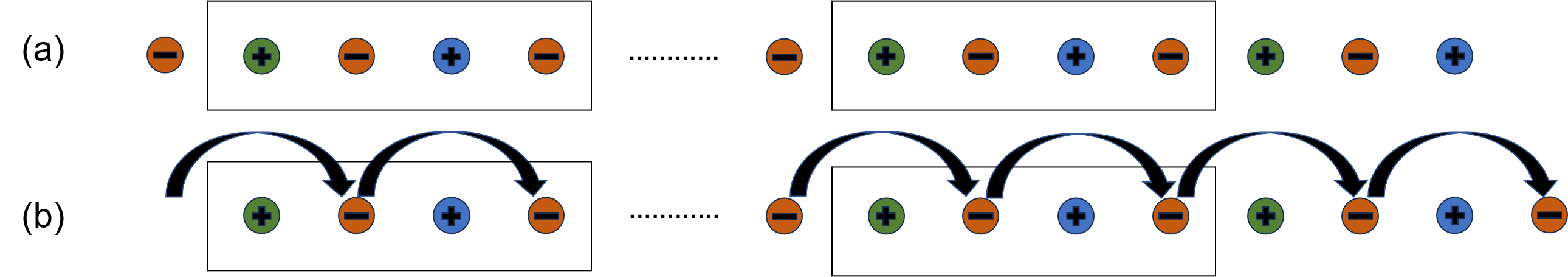}
	\caption{\label{fig:2+} A one-dimensional chain with the black square representing the unit cell. When the anion moves by half a lattice vector along the black arrow, the chain changes from (a) to (b).}
\end{figure}

When considering the change in polarization $\Delta P$ during ferroelectric switching, according to Eq.~\ref{e0},
\begin{align}\label{e2}
	\Delta \mathbf{ P}
	&=\Delta\mathbf{ P_{U}}+\Delta\mathbf{ P_S}.
\end{align}
And according to Eq.~\ref{ep} under PBC, we can obtain another form
\begin{align}\label{e1}
	\Delta \mathbf{ P}
	&=\Delta\mathbf{ P_{U}}+\Delta m\mathbf{ P_q}.
\end{align}
The latter term in Eq.~\ref{e1} arises from the additional indeterminacy introduced by the PBC assumption, and in order to obtain the correct $\Delta P$, $\Delta m$ must be determined via the specific surface variation $\Delta P_S$. Eq.~\ref{e2} implies that we need to consider the $\Delta P_S$ induced by surface charge changes\cite{PhysRevB.9.1998}.
In conventional ferroelectric materials, ionic displacements during the ferroelectric phase transition usually occur within a single unit cell, so that $\Delta P_S = 0$. Therefore, neglecting the surface contribution does not cause serious issues in calculating $\Delta P$\cite{Resta2007}.  However, for cases where the surface conditions change before and after a ferroelectric phase transition, this must be properly considered. 
As an example, when we move all the anion by half a lattice unit in the 1D chain, from  Fig.~\ref{fig:2+}a to  Fig.~\ref{fig:2+}b,
\begin{align}
	&\Delta \mathbf{ P_{U}}=0,
	\\
    &\Delta \mathbf{ P_{S}}=-1.
\end{align}
This demonstrate that both \(P_S\) and \(\Delta P_S\) cannot be neglected in general.

\begin{figure}[!t]
	\includegraphics[width=8cm]{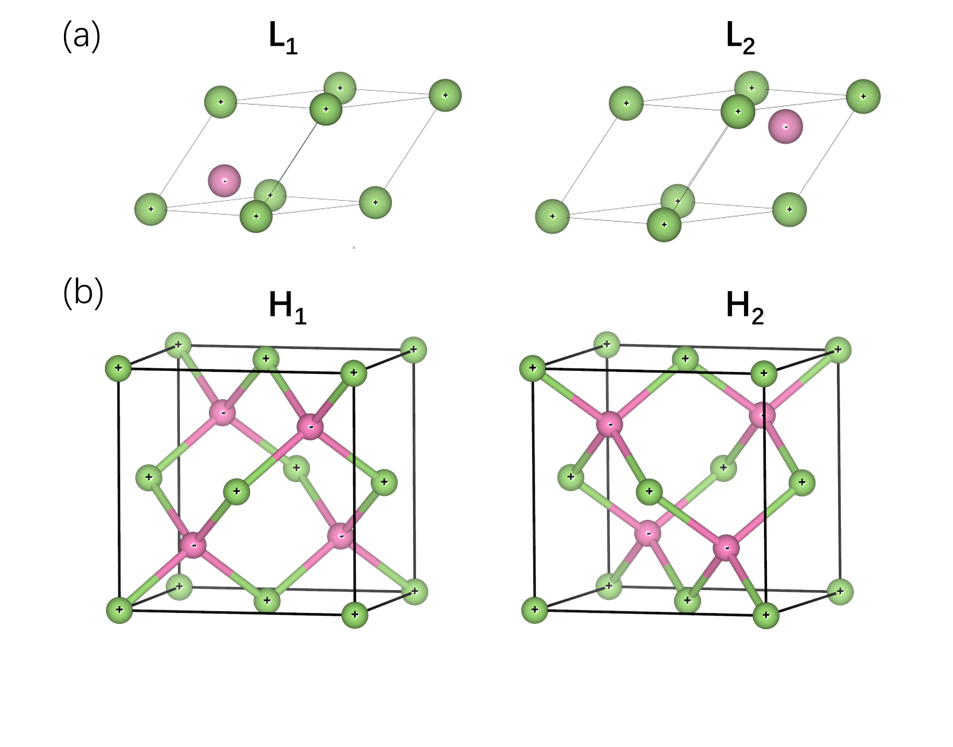}
	\caption{\label{fig:3} (a) The primitive cell \(L\) of the zinc-blende structure. The configuration on the left is denoted as \(L_1\). By moving the anion along the (111) direction by \((1/2, 1/2, 1/2)\), we obtain the configuration on the right, denoted as \(L_2\). (b) The conventional cell \(H\) of the zinc-blende structure. The left configuration is denoted as \(H_1\) and the right as \(H_2\). In the figure, green spheres represent the positive charge centers, and pink spheres represent the negative charge centers.
	}
\end{figure}

We now apply the foregoing mathematical reasoning to one of the most classic crystal structures—the zinc-blende (ZB) structure—which is normally considered as non-polar material with \( T_d \) symmetry when PBC is applied.
As shown in Fig.~\ref{fig:3}, we can view the ZB crystal with primitive unit cell \(L\) and conventional unit cell \(H\).
In Fig.~\ref{fig:3}a, the primitive ZB cell consists of one positive charge and one negative charge. The movement from configuration \(L_1\) to configuration \(L_2\)  results in a change in polarization \(\Delta P\)

\begin{align}
	\Delta \mathbf{P_{U_L}}=& -\frac{1}{V_0}[(3/4,3/4,3/4)-(1/4,1/4,1/4)]
	\\
	=& -\frac{1}{V_0}(1/2,1/2,1/2). \notag
\end{align}
Where \( V_0 \) represents the volume of the primitive cell.
However, in Fig.~\ref{fig:3}b, the change in polarization \( \Delta P_{U_H} \) between configurations \(H_1\) and \(H_2\) is given by:
\begin{align}
	\Delta \mathbf{P_{U_H}}= -\frac{1}{4V_0}[&(1/4,1/4,3/4)-(1/4,1/4,1/4)
	\\
	+&(1/4,3/4,1/4)-(1/4,3/4,3/4)\notag
\\
+&(3/4,1/4,1/4)-(3/4,1/4,3/4)\notag
\\
+&(3/4,3/4,3/4)-(3/4,3/4,1/4)]\notag
\\
	= 0.~~~~~~& \notag
\end{align} 

Note that when PBC is applied to both cases in Fig.~\ref{fig:3}, \(L_1\) is equivalent to \(H_1\) and \(L_2\) is equivalent to \(H_2\). This implies that neglecting the surfaces prevents the unique determination of \(\Delta P\) from the initial to final states, further suggesting that PBC are unsuitable for this system.

The origin of the above observed polarization in a  'non-polar' system become clear if we treat the above example as macroscopic crystals and take into account the contribution of surface. The unit cell choices in Fig.~\ref{fig:3}a and \ref{fig:3}b correspond to different macroscopic crystals when surface is taking into account. For macroscopic crystals, the corresponding symmetries of \( L_1 \) and \( L_2 \) are \( C_{3v} \), which is polar and can have non-zero polarization, while \( H_1 \) and \( H_2 \) are \( T_d \), a non-polar point group, which requires \( P_{U_{H_1}} = P_{U_{H_2}} = 0 \). Imposing a 3D periodicity can lead to the misconception that \( L \) and \( H \) are identical \cite{ji2024fractional}, both exhibiting \( T_d \) symmetry, which in turn leads to the unphysical conclusion that non-polar crystals can have polarization and be ferroelectric.

\begin{figure}[!t]
	\includegraphics[width=8cm]{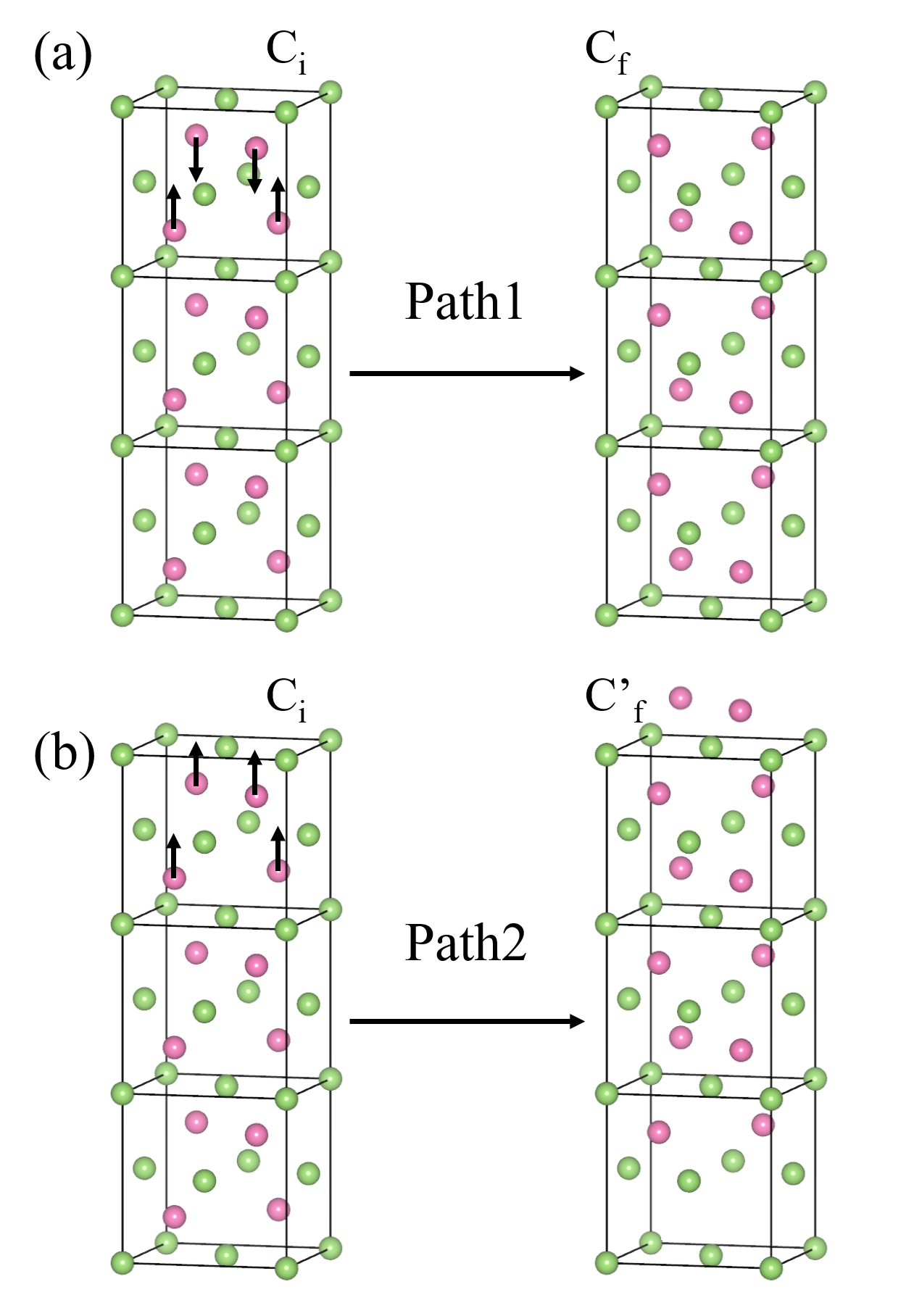}
	\caption{\label{fig:4} Two different phase transition paths from \( H_1 \) to \( H_2 \).The black arrows indicate the direction of anion movement. \( C_i \) represents the macroscopic crystal corresponding to \( H_1 \). (a) The final state obtained from \( C_i \) moving along \(Path1\) is \( C_f \); (b) the final state obtained from \( C_i \) moving along \(Path2\) is \( C'_f \).
	}
\end{figure}

To further demonstrate the contribution from the surfaces, we show in Fig.~\ref{fig:4} two different phase transition pathways from \( H_1 \) to \( H_2 \).
\begin{align}
	\mathbf{P_{C_i}}
	&=\mathbf{P_{U_{H_1}}}
	\\
	&=0, \notag
	\\
	\mathbf{P_{C_f}}
	&=\mathbf{P_{U_{H_2}}}
	\\
	&=0, \notag
	\\
	\mathbf{P_{C'_f}}
	&=\mathbf{P_{U_{H_2}}} + \mathbf{P_S}
	\\
	&=0+\frac{-1}{2V_0}(0,0,1) \notag
	\\
	&=-\frac{1}{2V_0}(0,0,1),   \notag
\end{align}
then,
\begin{align}
	\Delta \mathbf{P_{Path1}}
	&=\mathbf{P_{C_f}}-\mathbf{P_{C_i}}
	\\
	&=0, \notag
	\\
	\Delta \mathbf{P_{Path2}}
    &=\mathbf{P_{C'_f}}-\mathbf{P_{C_i}}
	\\
	&=-\frac{1}{2V_0}(0,0,1).   \notag
\end{align}
For periodic systems, we cannot distinguish between \( C_f \) and \( C'_f \); However, for macroscopic systems, \( C_f \) and \( C'_f \) have distinct differences: \( C_f \) is entirely composed of \( H_2 \), while \( C'_f \) has a new surface \( S \) due to the overall upward shift of the anions. It is this newly formed surface that causes \( \Delta P_{Path2} \) to be non-zero. In fact, compared to the small changes in \( \Delta P \) that occur during traditional ferroelectric phase transitions, the large fractional \( \Delta P \) in the ZB structure corresponding to the space group \( F\overline{4}3m \) results from the movement of atoms from the 4c to the 4d Wyckoff positions. The underlying physical reason is that the movement of atoms between Wyckoff positions leads to a change in the surfaces, which causes a large \( \Delta P_S \)~\cite{wang2022unconventional,wumenghao1}. This implies that crystals forced into non-polar groups by imposed PBC, and thus regarded as incapable of ferroelectric switching, actually exhibit polar symmetry once surfaces are considered and represent potential new types of ferroelectric materials\cite{ji2024fractional,XiangPhysRevLett.134.016801}.

Based on the above discussion, we present the following general conclusions: (i) The polarization \( P \) of a macroscopic crystal consists of two parts: \( P_U \) and \( P_S \). Regardless of the number of repeating cells in the solid, both \( P_U \) and \( P_S \) are equally important and in general cannot be neglected. (ii) Although \( P_U \) and \( P_S \) depend on the choice of repeating unit cell, the polarization \( P = P_U + P_S \) of a macroscopic crystal is independent of the choice of unit cell and is a well-defined and single-valued quantity. (iii)  ~\( \Delta P_U \) and \( \Delta P_S \) are equally important, and \( \Delta P_S \) is not always a negligible quantity. Furthermore, the above discussions suggest that (iv) for crystals of any space group, by properly choosing the unit cell and displacement path, and repeat the unit cell, one could, in principle, tailoring the combined bulk-plus-surface system belong to a polar point group, thereby enabling a broad new class of materials with polarization or even being ferroelectric.

It should be emphasized that our above analysis and arguments concerning the physical properties of real systems do not imply that the MTP is wrong in calculating ferroelectric polarization. On the contrary, the MTP and our discussion are not in conflict but are fully consistent. When we actually calculate the polarization of a real solid, it is difficult to compute \(P_S\) directly from Eq.~\ref{q1} because of the complexity of effects such as surface reconstruction. However, according to the surface theorem~\cite{vanderbilt1993,JCPresta,RMPBOYS}, one can show that the contribution from the finite displacements of ions and the electronic cloud induced by surface reconstruction vanishes in the macroscopic limit (\(n \to \infty\)). This implies that the magnitude of \(P_S\) defined in this work is strictly equal to \(m P_q\), where the integer \(m\) is completely determined by the net charge on a single surface. For example, if the left surface carries a charge of \(5e\), then, due to overall charge neutrality, the right surface must carry \(-5e\), and thus \(m = 5\). This means that we only need to count the static charge on the chosen surface in order to obtain the exact value of \(P_S\).

For the unit polarization \(P_U\), we can follow the standard procedure of the modern theory of polarization (MTP) by first idealize the real crystal as an infinite periodic system. Using first-principles band-structure calculations with PBC, the usual output is the cell-periodic electronic Bloch function \(|u_{n\mathbf{k}}\rangle\). It is then convenient to compute the unit polarization \(P_U\) using the Berry-phase / Wannier-center formulation~\cite{berry1984,zak1989,kingsmith1994,resta1994,martin2004,marzari1997} as follows:

\begin{align}
	&\mathbf{P_U} = \frac{1}{V} \left( \sum_i \left( q_i \mathbf{r}_i \right)^{\text{ions}} + \sum_n^{\text{occ}} \left( q_n \mathbf{\bar{r}_n} \right)^{\text{WFs}} \right).
	\\
	&\mathbf{\bar{r}_n} = i \frac{\Omega}{(2\pi)^3} \int_{\text{BZ}} d^3k \, e^{-i\mathbf{k} \cdot \mathbf{R}} 
	\left\langle u_{nk} \left| \frac{\partial u_{nk}}{\partial \mathbf{k}} \right. \right\rangle.
\end{align}

Because of the multivalued nature of \(\bar{{r}}_n\) under PBC, the \({P}_U\) obtained from the above formulas is likewise multivalued. The next step is to “cut” the real system of interest out of this idealized periodic crystal, i.e., to introduce surfaces. It is crucial at this stage to choose surfaces that are consistent with those of the actual system. This requirement also removes the arbitrariness in the choice of unit cell, so that \({P}_U\) is no longer multivalued but collapses onto a specific branch of the polarization lattice determined by the chosen surfaces. After carrying out these steps, we obtain the polarization of the real macroscopic crystal. We emphasize once again that, although the polarization of the purely periodic PBC system is multivalued, the polarization of the real, cut-out system, \({P} = {P}_U + {P}_S\), is single valued and well defined.

In summary, we have used one-dimensional chains and zinc-blende structures as examples to show the importance of surface contribution to the description of  polarization of macroscopic crystals. Neither \( P_S \) nor \( \Delta P_S \) can be neglected no matter how large the crystal size is. This indicate that in general, PBC, which ignore the true surface contribution, is not sufficient to describe polarization of a real crystal. To say in another words, a macroscopic crystal is not a periodic crystal when the polarization is concerned. We have shown that with the correct treatment of surface, the polarization is independent of the choice of the unit cell and is single valued for a specific surface. For macroscopic crystals, symmetry classification should account for both bulk and surface conditions. The symmetry of polarized crystals must still belong to polar point groups, while the polarization of crystals belonging to non-polar point groups must be zero. Crystals that were traditionally misclassified as non-polar under imposed PBC actually exhibit polar symmetry once surfaces are considered, thereby revealing their potential to become new types of ferroelectric materials. These insights not only reveal the inapplicability of PBC to non-local physical quantities such as polarization, but also highlight the important role of surfaces in ferroelectric phenomena, and further make it possible to realize a new class of ferroelectric materials through surface-engineered switching.

This work is supported by the National Key Research and Development Program of China (Grants No. 2024YFA1409800) and the National Natural Science Foundation of China (Grants No. 12088101).

\bibliography{ku}

\end{document}